\documentclass[apj,twocolappendix,numberedappendix]{emulateapj}

\usepackage{savesym}
\usepackage{natbib}

\usepackage{amsmath,amssymb,amsthm,mathrsfs,dsfont}
\usepackage{graphicx}
\usepackage{subfigure}
\usepackage{captcont}
\usepackage{float}
\usepackage{booktabs}
\usepackage{epstopdf}
\usepackage{color}

\def\be{\begin{equation}}
\def\ee{\end{equation}}
\def\kms{{\rm \,km\,s^{-1}}}

\def\Gyr{{\rm \,Gyr}}
\def\e{{\rm e}}
\def\Mpc{{\rm \,Mpc}}
\def\kpc{{\rm \,kpc}}

\def\msun{{\,M_\odot}}

\begin{document}

\title{A Baryonic Effect on the Merger Timescale of Galaxy Clusters}
%
%
\shortauthors{Zhang, Yu, \& Lu}
\author{Congyao Zhang$^1$, Qingjuan Yu$^{1,\dagger}$, and Youjun Lu$^2$}
\affil{
$^1$~Kavli Institute for Astronomy and Astrophysics, Peking
University, Beijing, 100871, China; $^\dagger$yuqj@pku.edu.cn \\
$^2$~National Astronomical Observatories, Chinese Academy of
Sciences, Beijing, 100012, China
}

\begin{abstract}
Accurate estimation of the merger timescale of galaxy clusters is important to
understand the cluster merger process and further the formation and evolution
of the large-scale structure of the universe. In this paper, we explore a 
baryonic effect on the merger timescale of galaxy clusters by using hydrodynamical
simulations. We find that the baryons play an important role in accelerating
the merger process. The merger timescale decreases with increasing the gas
fraction of galaxy clusters. For example, the merger timescale is shortened by
a factor of up to 3 for merging clusters with gas fractions $0.15$, compared
with the timescale obtained with zero gas fractions. The baryonic effect is
significant for a wide range of merger parameters and especially
more significant for nearly head-on mergers and high merging velocities.
The baryonic effect on the merger timescale of galaxy clusters is expected to
have impacts on the structure formation in the universe, such as the cluster
mass function and massive substructures in galaxy clusters, and a bias
of ``no-gas'' may exist in the results obtained from the dark
matter-only cosmological simulations.
\end{abstract}


\section{Introduction} \label{sec:introduction}

Groups and clusters of galaxies are assembled by mergers and accretion of
small structures (e.g., dark matter halos and galaxies) in the $\Lambda$ Cold
Dark Matter (CDM) cosmology \citep[e.g.,][]{Kravtsov2012}. Mergers play a
central role in the structure formation processes and have been extensively
studied through numerical simulations (\citealt{LC94,Navarro1995,Jenkins2001,Fakhouri2010,Genel08};
for analytical models and observations, see \citealt{Lacey1993,Okabe2008}).
Most of the simulations that address the statistical properties of the mergers
(e.g., the merger rate, the cluster mass function, and substructures of groups
and clusters) are dark matter-only simulations and the gastrophysics of baryonic
materials is ignored. Ignoring the gastrophysics is at least due to the
following reasons: (1) dark matter (DM) dominates the gravitational field;
and (2) it is computationally difficult for hydrodynamical simulations to
reach sufficiently high spatial resolutions and statistical precisions
simultaneously. However, the negligibility of the baryonic effects, particularly
on the scale of galaxy clusters ($M\ga 10^{14}\msun$), is not always self-evident
\citep{Stanek2009,Cui2012}. In some simulations, gastrophysics has been
implemented to understand the behaviors of baryonic materials in the mergers
\citep[e.g.,][]{Poole2006,Zuhone2011}.

In this work, we show that the baryonic materials\footnote{In this paper, the
baryonic materials mainly refer to the hot gas in the DM halos of galaxy
clusters, and stars and cold gas in galaxies are not taken into account due
to their low mass fractions ($\sim2\%$ of the total matter; see \citealt{Dai2010,Lagana2011,Wu2015})
in the galaxy clusters. 
}
play an important role in the merger of two clusters and help to accelerate the
merging process, which suggests that a bias of ``no-gas'' exists in the current
DM-only cosmological simulations (hereafter denoted as the ``no-gas'' bias).
Note that in the numerical studies on the merger timescale of galaxies by
\citet{Jiang2008,Jiang2010} and \citet{Kolchin2008}, dynamical friction
among the collisionless (i.e., DM or stars) particles plays the dominant
role in the energy and angular momentum dissipation during the merger.
The hot gas in the DM halo of those merging galaxies has hardly any effect
due to its little amount (see the observational baryon fractions
in systems with different scales in \citealt{Dai2010} and references therein).
However, the fraction of baryonic matters to the total mass in clusters
can be significantly larger, approaching to the cosmic mean value
($\simeq 0.17$; see \citealt{Battaglia2013,Mantz2014}).  The effect of
the dissipative motion of baryons (unlike that of the collisionless DM
particles) on the merger timescale of galaxy clusters, is still poorly studied.
In this work, we perform a series of hydrodynamical (and DM-only) simulations
of mergers of galaxy clusters to single out the baryonic effect on the merger
processes. We find that the merger processes are significantly accelerated in
the hydrodynamical simulations and the merger timescale can be shortened by a
factor of up to 3, compared with those obtained from the
DM-only simulations.

This paper is organized as follows. In Section~\ref{sec:method}, we describe
our numerical simulations of mergers of two galaxy clusters and the method to
estimate their merger timescales. In Section~\ref{sec:result}, we present our
simulation results of the baryonic effect on the merger timescale and its
dependence on the physical parameters of the mergers. We also develop a toy
model to quantitatively interpret those results obtained from the simulations.
In Section~\ref{sec:discuss}, we discuss some possible implications of the
shortening of the merger timescales caused by the baryonic effect.
Finally, conclusions are summarized in Section~\ref{sec:conclusion}.

\section{Method} \label{sec:method}
\subsection{The Numerical Simulations} \label{sec:method:simulation}

We perform numerical simulations for the mergers between two galaxy
clusters, by using the smoothed particle hydrodynamics (SPH) code, GADGET-2
\citep{Springel2001na, Springel2005}. The galaxy clusters in the simulations
are simplified as the spherical halos consisting of collisionless DM particles
and collisional gas particles.
The gas mass fraction of galaxy clusters is defined by the ratio
of the gas mass to the total mass within the virial radius.
The initial gas density profile is assumed to follow the Burkert profile
\citep{Burkert1995}, and is normalized according to the assumed gas fraction.
The similar runs are described in
\citet{Zhang2014,Zhang2015} (see detailed settings of the initial conditions therein).
The gas is assumed to be adiabatic in the simulations, i.e., neither radiative
cooling nor energetic feedback is involved. The masses of the gas and the DM
particles are $7.5\times10^{7}\msun$ and $4.2\times10^{8}\msun$,
respectively. We have chosen other different mass resolutions (e.g., two times
lower or five times higher than the above settings) to examine the convergence
of our simulation results and found that all the results related to the purpose
of this work are converged.

The settings of the initial merger parameters, i.e., the masses of the primary
and the secondary clusters $M_1$ and $M_2$ ($M_1\geq M_2$), their mass ratio $\xi\
(\equiv M_1/M_2)$, gas fraction $f_{\rm gas}$, initial relative velocity $V$,
and the impact parameter $P$, are summarized in Table~\ref{tab:ic_para}. For each
parameter set ($M_1,\,\xi,\,V,\,P$), we choose several different gas fractions
to explore the dependence of the merger timescales on the gas fraction.

\begin{table*} [htb!]
\begin{center}
\caption{Initial merger parameters}
 \label{tab:ic_para}
\begin{tabular}{c|c|c|c|c}
  \hline
\hline
 $M_1\ (10^{14}\msun)$ & $\xi$ & $f_{\rm gas}$ & $V\ (\kms)$ & $P\ (\kpc)$ \\
  \hline
  $1.0$ & 2 & 0.0, 0.15 & 500 & 0 \\
  \hline
  $2.0$ & 1, 2, 4 & 0.0, 0.05, 0.1, 0.15 & 250, 500, 750, 1000 & 0 \\
  \hline
  $2.0$ & 1, 2 & 0.0, 0.05, 0.1, 0.15 & 500, 1000 & 100, 200, 500, 1000 \\
  \hline
  $5.0$ & 1, 2, 4 & 0.0, 0.15 & 500, 1000 & 0 \\
  \hline
\hline
\end{tabular}
\end{center}
\end{table*}

\subsection{The Merger Timescales} \label{sec:method:timescale}

\begin{figure}
\centering
\includegraphics[width=0.45\textwidth]{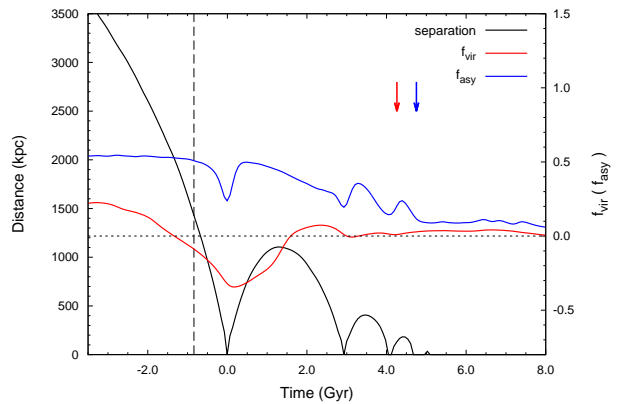}
\caption{Evolution of some physical properties of two merging clusters,
including the separation between their mass centers (black), the virial factor
(red; Eq.~\ref{eq:fvir}), and the asymmetry index of the merging system
(blue; Eq.~\ref{eq:fasy}). The initial condition of the merger is
$(M_1,\,\xi,\,V,\,P,\,f_{\rm gas})=(2\times10^{14}\msun,\,2,\,250\kms,\,0,\,0)$.
The vertical dashed line indicates the crossing time
$t_{\rm cross}\,(=-0.84\Gyr)$, the moment when the center of the
secondary cluster first crosses the virial radius of the primary
one. The red and the blue arrows indicate the virialization moments
$t_{\rm relax,\,vir}\,(=4.26\Gyr)$ and $t_{\rm relax,\,asy}\,(=4.75\Gyr)$,
respectively. The merger timescale of this system is $5.59\Gyr$.
See Section~\ref{sec:method:timescale}.}
\label{fig:example}
\end{figure}

The merger timescale, denoted by $\Delta t$, can be estimated from the
snapshots of the simulations, which is defined as the time duration between the
moment when the center of the secondary cluster first crosses the virial radius
of the primary one ($t_{\rm cross}$) and the moment when the system finally
reaches the relaxation state ($t_{\rm relax}$; see Eq.~\ref{eq:trelax} below),
i.e.,
\be
\Delta t \equiv t_{\rm relax} - t_{\rm cross}.
\label{eq:timescale}
\ee
The relaxation time $t_{\rm relax}$ can be identified in the following
approaches based on either the time evolution of virial factor or the asymmetry
index, which reveals the evolution of the dynamical state or the morphology of
the merger system.
\begin{enumerate}
\item The virial factor ($f_{\rm vir}$) is defined by
\be
f_{\rm vir} \equiv 1+\frac{2T}{U+S},
\label{eq:fvir}
\ee
where $T$ and $U$ are the kinetic and the potential energy of the merger
system, respectively, and $S$ is a surface pressure term (Equations~(5.126)--(5.129)
in \citealt{Mo2010}; see also \citealt{Poole2006,Ricker1998}).
\citet{Poole2006} show that the virial factors for the gas and for the DM
track similar trends, although the oscillating evolution of the gas virial
factor damps faster than that of the DM virial factor.  In this work, the
virial factor is calculated from the DM particles located within the radius
$r_{\rm vir}$ (where $r_{\rm vir}$ is defined as the virial radius of the
galaxy cluster with mass $M_1+M_2$)\footnote{We have checked that the virial factor is
not sensitive to the value of the selected radius if it is in the range of
$0.8-1.2r_{\rm vir}$.}.  The moment for the ``completion'' of the virialization
process ($t_{\rm relax,\, vir}$) is defined as the time point after which the
system maintains $|f_{\rm vir}|<5\%$ and the variance of $f_{\rm vir}$ has been
smaller than $1\%$ continuously for more than $1.5\Gyr$.
\item The asymmetry index of the merger system ($f_{\rm asy}$) is defined by
\be
f_{\rm asy} \equiv \sqrt{\frac{\Sigma(I_0-I_\pi)^2}{2\Sigma I_0^2}},
\label{eq:fasy}
\ee
where $I_0$ is the mass surface density distribution of the original merger
system, and $I_\pi$ is the merger system transformed by rotating the original
one with $180\arcdeg$ around the position of the maximum of the surface mass
density. The summation $\Sigma$ includes all the pixels of the 2-dimensional
surface density image (see \citealt{Conselice2000}, where a similar variable
is used to quantify galaxy morphologies). The virialization moment
($t_{\rm relax,\,asy}$) is defined as the time point when the system starts
to maintain $f_{\rm asy}<15\%$.
\end{enumerate}
Finally, we define the relaxation moment of the merger system in this study as
\be t_{\rm relax}\equiv \max(t_{\rm relax,\,vir},\,t_{\rm relax,\,asy}).
\label{eq:trelax}
\ee
As an example, Figure~\ref{fig:example} shows the evolution of the separation between
two merging clusters and the corresponding evolutions of the virial factor
and the asymmetry index of the merging system. We find that in most of our simulations,
the relaxation moment is close to the moment of the third or fourth pericentric passages,
which can be seen from the example case illustrated in Figure~\ref{fig:example}.

\section{Results} \label{sec:result}
\subsection{The Baryonic Effect on the Merger Timescales} \label{sec:result:merger}

\begin{figure}
\centering
\includegraphics[width=0.45\textwidth]{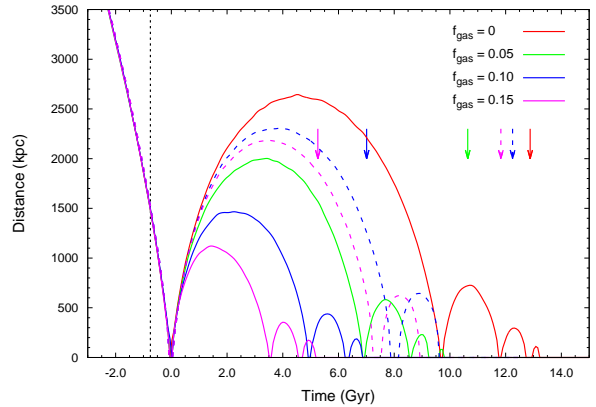}
\caption{Evolution of the separation between two cluster mass centers for the
merger with different gas fractions but fixed parameters
$(M_1,\,\xi,\,V,\,P)=(2\times10^{14}\msun,\,2,\,1000\kms,\,0)$. The solid lines
represent those mergers in which the gas fractions of the primary and secondary
clusters are the same with $0$ (red), $0.05$ (green), $0.10$ (blue), and
$0.15$ (magenta), respectively. The dashed lines represent the mergers in which
the gas fraction of the secondary cluster is zero and that of the primary
cluster is 0.15 (blue) and 0.225 (magenta), respectively (the rest merger
parameters are the same as those with the solid lines). The vertical dotted
line shows the crossing time $t_{\rm cross}$ and the arrows indicate the
relaxation moments $t_{\rm relax}$ for the corresponding mergers with the same
line colors and line styles (see Eq.~\ref{eq:timescale}).  This figure shows
that the presence of dissipative baryons in the mergers of galaxy clusters can
lead to a significant decrease of the apocenter distances and an acceleration
of the merger processes. See details in Section~\ref{sec:result:merger}.
}
\label{fig:timescale}
\end{figure}

Figure~\ref{fig:timescale} shows the evolution of the separation between the mass
centers of the two merging clusters, for mergers with different gas fractions.
The solid lines represent those mergers of which both the gas fractions of the
two progenitor clusters are 0 (red), 0.05 (green), 0.10 (blue), and 0.15
(magenta), respectively.  As seen from the solid lines, the first apocenter
distance decreases with increasing $f_{\rm gas}$. The first apocenter distance
for the case with $f_{\rm gas}=0.15$ is smaller than that for the case with
$f_{\rm gas}=0$ by a factor of $\sim 2-3$, and correspondingly the time
duration between the primary and the secondary pericentric passages is smaller
by about a similar factor.  The arrows in the figure indicate the relaxation
moments of the mergers.  For the DM-only case ($f_{\rm gas}=0$), the merger
timescale can be as long as $13.6\Gyr$, close to the age of the universe.
However, for the case with $f_{\rm gas}=0.15$, the timescale is about
$6.0\Gyr$, approximately half of the DM-only one. Thus, the higher gas fraction
produces a shorter merger timescale.  The figure suggests that the baryons have
a remarkable effect on the merger processes and they help to accelerate the
virialization of the merging system.

The baryonic effect on the merger timescale can be understood in the following
way. The collisionless DM halos pass through each other after the pericentric
passage; however, the collisional gas does not. The gas pressure impedes the gas
from moving with its DM hosts and leaves the gas to be around the mass center
of the entire system. The gas contributes an additional gravitational force
from the center, which drags the runaway DM particles to fall back more quickly
and makes them experience more dynamical friction at the relatively early stage.
Though the gas mass is only around one tenth of the total cluster
mass, the distance between the gas mass center and the DM halo is
approximately half of the distance between the DM haloes. Since the gravity is
inversely proportional to the
square of the distance, the gravitational force contributed by the gas mass can
be significant compared with that contributed by the DM halo mass.

We run two additional simulations (shown as the dashed lines in
Figure~\ref{fig:timescale}) to support the above scenario. These two
simulations have the same initial merger parameters $(M_1, \xi, V, P)$ as the
others shown in Figure~\ref{fig:timescale}, but the gas fraction of the second
cluster is zero and that of the primary cluster is 0.15 (blue dashed line) or
0.225 (magenta dashed line).  The gas fractions of the entire merger systems
for the above two simulations are $0.10$ and $0.15$, the same as those two
shown by the solid blue and the solid magenta lines, respectively.  However,
there is no collision of two gas halos in those two simulations.  As seen from
Figure~\ref{fig:timescale}, the mergers in which the gas halos collide have
much shorter merger timescales than those without gas halo collisions, which
implies that the additional gravitational potential contributed by the gas left
around the center of mass of the entire system after the pericentric passage
plays the dominant role in the baryonic effect on the merger timescale.
Hereafter in this work the two merging clusters are assumed to have a same gas
fraction.

A toy model is developed in Section~\ref{sec:result:toymodel} below to
describe the above scenario quantitatively.

\subsection{Dependence of the Merger Timescales on the Physical Parameters of
the Merger Systems } \label{sec:result:parameter}

\begin{figure*}
\centering
\includegraphics[width=0.8\textwidth]{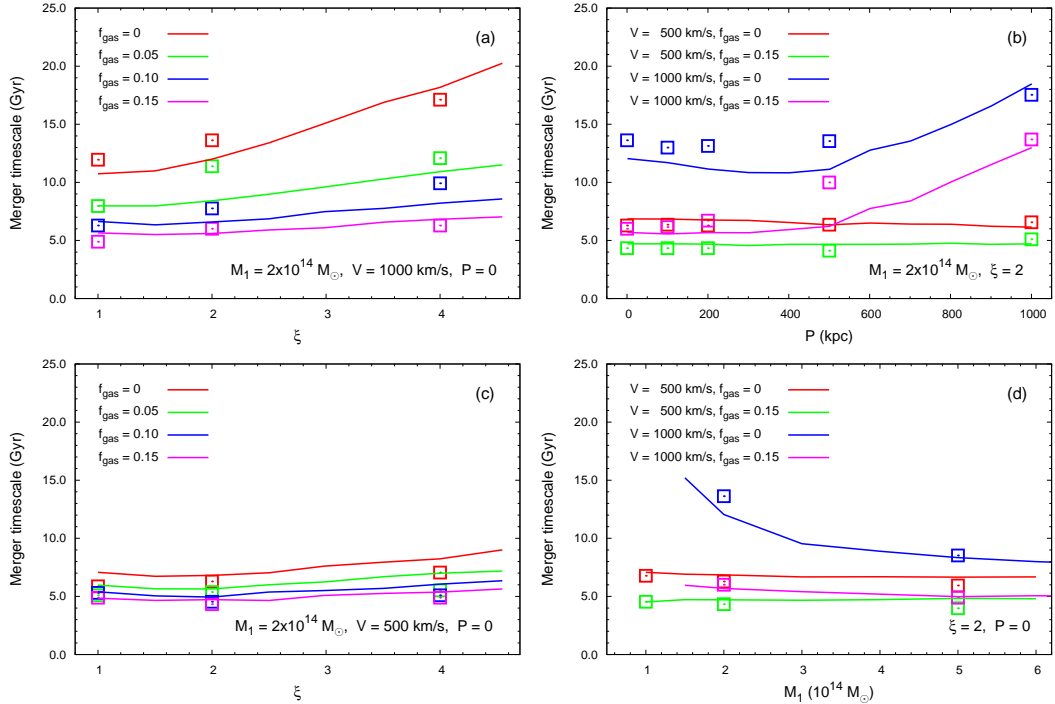}
\caption{Dependence of the merger timescale on different merger parameters.
Panels (a) and (c) show the dependence on the mass ratio $\xi$ for mergers with
initial relative velocities $V=1000\kms$ and $500\kms$, respectively. Panels
(b) and (d) show the dependence on the impact parameter and the mass of the
primary cluster, respectively.  The points in each panel represent the
timescales measured from the numerical simulations (see
Section~\ref{sec:result:parameter}), and the solid lines represent the
timescales obtained from the toy model (see Section~\ref{sec:result:toymodel}).
In panel (d), no data points are shown for the cases with $V=1000\kms$ at
$M_1=10^{14}\msun$, because the merger timescales of these systems are much longer
than the Hubble timescale (similarly in Fig.~\ref{fig:ratio}c below).
This figure shows that the baryonic effect accelerates the virialization
processes for merger systems with a wide range of the merger parameters.}
\label{fig:parameter}
\end{figure*}

Figure~\ref{fig:parameter} shows the dependence of the merger timescale on the
physical parameters of the merger systems. The points represent the timescales
measured from the numerical simulations, and the solid lines represent the
estimates obtained from the toy model (described in
Section~\ref{sec:result:toymodel} below). Panels (a) and (c) of the figure show
the merger timescale as a function of mass ratios for those mergers with
initial relative velocities $V=1000$ and $500\kms$, respectively. Panels (b)
and (d) show the merger timescale as a function of the impact parameter and the
mass of the primary cluster with different gas fractions, respectively. As seen
from the figure, the merger timescales for the mergers with $f_{\rm gas}=0.15$
are always shorter than those with $f_{\rm gas}=0$, given the same other
parameters. The figure suggests that the merger process is accelerated due to
the presence of the gas component for a wide range of physical parameters.

\begin{figure*}
\centering
\includegraphics[width=0.8\textwidth]{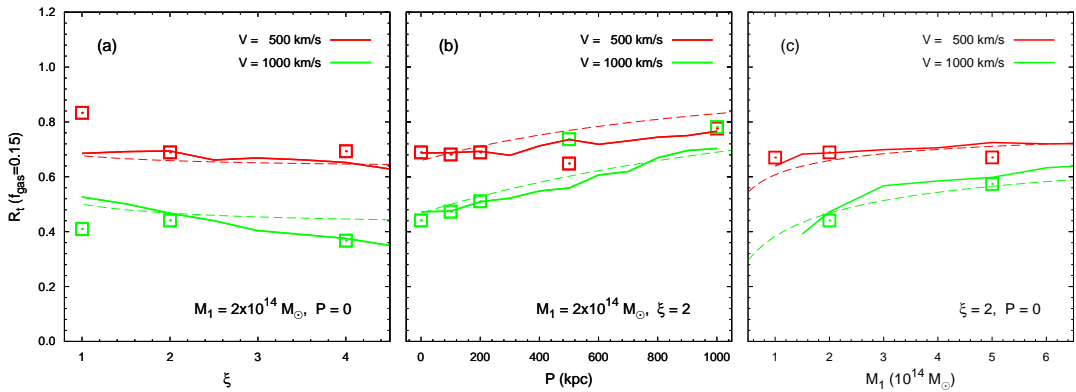}
\caption{The ratios of the merger timescales with
$f_{\rm gas}=0.15$ to those obtained with $f_{\rm gas}=0$. Panels (a)--(c)
show the dependence of the timescale ratios on the mass ratio, the impact
parameter, and the mass of the primary cluster for different relative velocities
(i.e., $V=500,\,1000\kms$), respectively. The points in each panel represent
the timescales measured from the numerical simulations, the solid lines represent the
timescales obtained from the toy model (see Section~\ref{sec:result:toymodel}), and the dashed lines are
the corresponding best-fit results from Equation~(\ref{eq:fitform}).
As seen from the figure, the timescale ratios are smaller than unity,
which indicates the acceleration of the virialization processes by the baryons.
The figure shows that the timescale ratio only mildly depends on
the parameters $\xi$ and $M_1$ (see Section~\ref{sec:result:parameter}).
}
\label{fig:ratio}
\end{figure*}

We use the ratio of the merger timescale ($R_{\rm t}$) estimated from the run
with $f_{\rm gas}=0.15$ to that with $f_{\rm gas} =0$ (i.e., DM-only simulations)
to quantify the baryonic effect revealed in Figure~\ref{fig:parameter}.
Figure~\ref{fig:ratio}(a)--(c) present the merger timescale ratios as a function
of mass ratios, impact parameters, and the mass of the primary cluster,
respectively. For each panels, we show the results for both $V=500\kms$ and
$V=1000\kms$. From the figure, we find the following main points.
\begin{enumerate}
\item The merger timescale ratios are smaller than 1, which indicates the
baryonic effect revealed in Figure~\ref{fig:parameter} above.
\item The merger timescale ratios obtained with $V=1000\kms$ are smaller than
those with $V=500\kms$ (except for the mergers with large impact parameters,
e.g., $P>500\kpc$), which suggests that the baryonic effect becomes more
significant for those stronger collisions with high relative velocities.
This dependence on $V$ can be understood as follows.
The strength of the dynamical friction drops significantly when
the separation between the two clusters gets larger, because the mass density is
low at the cluster outskirts (see Eq.~\ref{eq:dynfriction}). For a relatively
high initial relative
velocity, the additional gravitational drag contributed by the baryons plays a
more remarkable role in enhancing the efficiency of the dynamical friction by
shortening the apocenter distance and the orbital period, particularly at
the early merger stage.
\item The timescale ratio does not significantly correlate with the mass ratio
or the mass of the primary cluster for both the low and the high relative velocity
mergers (i.e., $V=500,\,1000\kms$; panels a and c). The baryonic effect on the
merger timescale only mildly depends on the merger parameters $\xi$ and $M_1$.
For the high relative velocity mergers, the timescale ratio shows a slightly
decreasing trend when $M_1\leq2\times10^{14}\msun$. Because the circular velocity
of the cluster with mass $2\times10^{14}\msun$  at the virial radius is
$\simeq800\kms$, the role of the baryons is more significant when $V=1000\kms$,
compared with that in the low-velocity mergers (due to the similar argument in the above item).
\item The dependence on the impact parameter (panel b) shows different behavior
for different $V$. The timescale ratio increases with increasing impact
parameter for mergers with $V=1000\kms$.  However, for the low-velocity mergers
(i.e., $V=500\kms$), the timescale ratio is roughly constant ($\sim0.7$) for
mergers with a large range of impact parameters from $0$ to $1000\kpc$. These
results can be understood in the scenario proposed in
Section~\ref{sec:result:merger} as follows.  The significance of the baryonic
effect on the merger timescale depends on the strength of the collision between
the gas halos and decreases with increasing primary pericenter distances of the
merger. For the low-velocity mergers, the primary pericenter distances are not
sensitive to the impact parameter. For example, for the mergers with
$V=500\kms\,(M_1=2\times10^{14}\msun,\,\xi=2$), the pericenter distances range
from 0 to $100\kpc$, when $P$ is from 0 to $1000\kpc$. However, when $V=1000\kms$
and $P=1000\kpc$ (with fixing the rest parameters), the pericenter distance
is up to $300\kpc$.
{\footnote {These primary pericenter distances estimated from the
simulations are generally consistent with those analytically calculated from the
two-body interacting systems (see Equations~(3)--(6) in \citealt{Khochfar2006}).}}
This explains the almost independence of the timescale ratio on the impact parameter
for the low-velocity mergers shown in Figure~\ref{fig:ratio}(b).
\end{enumerate}

According to the dependence behavior of the ratios of the merger timescales and
the gas fraction smaller than $0.2$, we fit our simulation results to the following form
(shown as the dashed lines in Figure~\ref{fig:ratio}),
\begin{eqnarray}
   &&R_{\rm t}(f_{\rm gas}) = \frac{1}{1+f_{\rm gas}\cdot X(M_1,\,\xi,\,V,\,P)},\quad {\rm and} \nonumber \\
   & & X(M_1,\,\xi,\,V,\,P) = \alpha\cdot\Big[1+\cfrac{V/10^3\kms}{((1+\xi^{-1})M_1/10^{14}\msun)^{1/3}}\Big]^{\beta}\cdot \nonumber \\
   & &\quad\quad\quad\quad\quad\quad\quad(1+P/10^3\kpc)^{\gamma},
\label{eq:fitform}
\end{eqnarray}
where $\alpha=1.24,\,\beta=3.44,\,\gamma=-1.34$.

Note that we mainly investigate the major mergers between galaxy clusters
($1\le\xi\le4$) in this study. For minor mergers (e.g., $\xi>10$), the baryonic
effect is not so significant, because (1) the tidal force and the ram pressure
is more effective in removing gas from the secondary cluster in minor mergers;
and (2) the gas fractions in galaxy groups are significantly smaller than those
in galaxy clusters \citep{Sun2012}.

\subsection{The Toy Model} \label{sec:result:toymodel}

We construct a simple toy model to quantitatively understand the scenario
proposed in Section~\ref{sec:result:merger}.
In the model, the two clusters are approximated as two point masses (or
particles) during the merger. The motion of each point mass is controlled by
the following three types of forces: the gravitational attraction from the
DM-halo of the other cluster ($\textbf{F}_{\rm halo}$; excluding the gas component, if
any); the dynamical friction ($\textbf{F}_{\rm df}$) due to the motion of one cluster in
the gravitational field of the other cluster; and the additional gravitational
force contributed by the baryons ($\textbf{F}_{\rm gd}$) dragged around the mass center
of the system after the primary pericentric passage. The total force
acting on each of the two particles (represented by the subscripts $i$)
can be written as
\be \textbf{F}_i=\textbf{F}_{{\rm halo},\,i}+\textbf{F}_{{\rm df},\,i}+\textbf{F}_{{\rm gd},\,i},\quad (i=1,\,2).
\label{eq:force} \ee
The baryonic effect discussed in Sections~\ref{sec:result:merger} and
\ref{sec:result:parameter} is included by the third term $\textbf{F}_{{\rm gd},\,i}$,
which is the main driver accelerating the merging process.
The orbits of each particle can be solved from the Newtonian equations of motion.
In this section, we summarize the main results obtained from the toy model and
compare them with those from the numerical simulations.  The detailed
implementation of the toy model and the calculation method are described
in Appendix~\ref{sec:appendix_A}.

The merger timescales estimated from the toy model are shown as the solid lines
in Figure~\ref{fig:parameter}, where the model adopts the same parameters
$(M_1,\,\xi,\,V,\,P,\,f_{\rm gas})$ used in the corresponding simulations. As
seen from Figure~\ref{fig:parameter}, the toy model reproduces the merger
timescales measured from the numerical simulations for most of the mergers and
also the dependence of the merger timescale on the physical parameters of the
mergers. This consistence not only indicates the robustness of the toy model,
but also implies the feasibility of the model to provide a simple and fast way
to estimate the merger timescale of galaxy clusters.  From this toy model, we
draw the following conclusions.
\begin{itemize}
\item The dynamical friction ($\textbf{F}_{\rm dyn}$) plays the dominant role in
shrinking the orbits of the merger system (see also Figure~\ref{fig:model} in
Appendix~\ref{sec:appendix_A}). The system cannot merge without the dynamical
friction term in Equation~(\ref{eq:force}).

\item The additional gravitational drag contributed by the baryons ($\textbf{F}_{\rm
gd}$) is responsible for the acceleration of the virialization process for
mergers with significant amount of gas. Without the gas drag term $\textbf{F}_{\rm gd}$
in Equation~(\ref{eq:force}), the toy model\footnote{Here, we adopt the gravitational force between the two clusters, instead of
the force between the two DM halos in Equation~(\ref{eq:force}) (i.e., $f_{\rm gas}=0$
in Eqs.~\ref{eq:gravforce} and \ref{eq:dynfriction}).}
cannot reproduce the dependence of
the merger timescale on gas fractions shown in Figure~\ref{fig:parameter}.
\end{itemize}

\section{Discussion} \label{sec:discuss}

We have demonstrated that the dissipative baryonic material can play an
important role in accelerating the merger processes of galaxy clusters in
Section~\ref{sec:result}.
The baryonic effect is relatively more significant for high relative velocity
mergers, compared with low relative velocity ones, as shown in
Figures~\ref{fig:parameter} and \ref{fig:ratio}.  Note that the pairwise
velocity distribution of DM halos ($>10^{14}\msun$) in cosmological simulations
shows its peak at $500-700\kms$ \citep{Thompson2012, Bouillot2014}, and the
fraction of the cluster mergers with high relative velocity $>750\kms$ can
be up to $\sim20\%$. Thus we expect that the baryonic effect is important
in the formation and evolution of massive clusters.  However, in most of the
current cosmological simulations (with comoving box size larger than a few
hundreds of Mpc), the dissipative nature of the gas component is not
considered, which may lead to the ``no-gas'' bias in those DM-only simulations.
In this section, we discuss the possible impacts of the baryonic effect on the
mass function of galaxy clusters at the high-mass end and the distribution of
the massive sub-structures in galaxy clusters as follows.
\begin{itemize}
\item First, the baryonic effect may have impacts on the cluster mass function
at the high-mass end. Since the major merger is an important channel for the
growth of galaxy clusters \citep{Fakhouri2008}, a massive cluster may be formed
earlier in a simulation including the baryonic effect, compared with that
formed in a simulation without including the baryonic effect.  As a
consequence, the high-mass end of the cluster mass function at a given redshift
might be higher than the expectation from the DM-only simulations.
\citet{Stanek2009} made a comparison between the cluster mass function
resulting from DM-only (or gravity-only) simulations and that from simulations
involving complex gas physics and found some difference
($\sim30\%$; see also \citealt{Cui2012,Cui2014,Cusworth2014,Martizzi2014,Bocquet2015}).
To extract a realistic cluster mass function from simulations,
we note that the following points need to be taken into consideration.
(1) The halo finding algorithm used in
simulations does not consider the dynamical state of halos, which may weaken
the baryonic effect revealed in this paper.
And (2) the box size of the simulations in \citet{Stanek2009}, \citet{Cusworth2014} (i.e., $500\,h^{-1}\Mpc$), and \citet{Cui2012,Cui2014} (i.e., $410\,h^{-1}\Mpc$)
may also limit the statistical precision at the high-mass end of the
mass function. A larger cosmological volume of hydrodynamical simulations
and more sophisticated mass function calibration are required
to investigate the possible ``no-gas'' bias in the cluster mass function.
\item The baryonic effect may have impacts on the spatial and mass distribution
of the massive substructures in galaxy clusters. The survival time for the
subclusters is suppressed by a factor of up to 3 in the mergers with
$f_{\rm gas}=0.15$, compared with the DM-only mergers (see the example
in Figure~\ref{fig:timescale}).  The subhalo mass function of the galaxy clusters
at the massive end ($m/M>0.1$, where $m$ and $M$ are the subhalo mass
and the host halo mass, respectively; see \citealt{Bosch2014}) might be
overestimated in the DM-only simulations. Thus the abundance of the massive
subhalos in galaxy clusters estimated from the simulations is expected to be
closer to the expectations from the analytical model in \citet{Bosch2014} (see
figs.~4 and 5 therein), after the baryonic effect is considered.
\end{itemize}

Note that the baryonic effect revealed in this paper is obtained under a
simplified model of the galaxy cluster structure and an assumption of adiabatic
gas dynamics. According to the results of the current cosmological simulations,
the effects of some other non-gravitational processes (e.g., radiative cooling,
energetic feedback) implemented in the simulations
\citep{Gnedin2004,Gnedin2011,Teyssier2011,Planelles2013} could be significant
on the gas fraction profile, but mild on the total gas fraction.  For example,
active galactic nuclei (AGN) heating may expel the gas out from the cluster
center \citep{Battaglia2013}, which might weaken the baryonic effect on the
merger timescale of galaxy clusters discussed in this paper.  To further
quantitatively understand the degree in affecting the acceleration of the
cluster mergers caused by the non-gravitational processes in realistic
structure formation, further analysis and simulations will be needed in the
future.

\section{Conclusion} \label{sec:conclusion}

In this paper, we present a baryonic effect on the merger timescale for the
mergers between galaxy clusters. A series of numerical simulations are
performed for mergers between two galaxy clusters with different settings of
the parameters ($M_1,\,\xi,\,V,\,P,\,f_{\rm gas}$). The baryons are found to
play an important role in accelerating the merger process. Our findings are
summarized as follows.
\begin{itemize}
\item The merger timescale of galaxy clusters strongly depends on their gas
fractions.  The merger timescales decreases with increasing the gas fraction.
We explore the parameter space for the mergers
and find that the baryonic effect is significant
for a wide range of merger parameters.  As an example, the merger timescale is
shortened by a factor of up to 3 for mergers with gas fraction 0.15,
compared with that with zero gas fraction (DM-only mergers; see
Figure~\ref{fig:ratio}). That baryonic effect can be understood as follows.
While the DM halos of the two clusters in the merger pass through each other
after the primary pericentric passage, their gas is however impeded at and
around the mass center of the entire merger system.  As a result, the central gas provides an
additional gravitational drag to the DM halos and the DM halos experience more dynamical friction
when passing through each other, which accelerates the merger process.
\item We explore the dependence of the baryonic effect on the merger parameters
($M_1,\,\xi,\,V,\,P$). We find the following points. (1) The above baryonic effect (i.e., the
dependence of the merger timescale on the gas fraction) depends on the initial
relative velocity; and generally, the effect is more remarkable if the relative
velocity is higher.  As an example, the merger timescales for mergers with
$f_{\rm gas}=0.15$ can be 2--3 times shorter than those DM-only mergers when
$(M_1,\,V,\,P)=(2\times10^{14}\msun,\,1000\kms,\,0)$. However, for the same
mergers but with $V=500\kms$, the merger timescales for $f_{\rm gas}=0.15$ are
only 1--2 times smaller than those of the DM-only cases. (2) The baryonic
effect does not significantly depend on the mass of the primary cluster and the
mass ratio. (3) The baryonic effect on the merger timescale is weakened in
the case with a large impact parameter (e.g., $P>400\kpc$), even if the
initial relative velocity is high (i.e., $V=1000\kms$).
\end{itemize}

The baryonic effect on the merger timescale of galaxy clusters may have impacts
on the structure formation in the universe, especially the cluster mass
function and massive substructures in galaxy clusters, which may result in the
``no-gas'' bias in the current DM-only cosmological simulations. It is
important to understand the ``no-gas'' bias and its implications in the future
studies.

This research was supported in part by the National Natural Science Foundation
of China under nos.\ 11273004, 11373031, and the Strategic Priority Research
Program ``The Emergence of Cosmological Structures'' of the Chinese Academy of
Sciences, Grant No. XDB09000000.

\appendix
\section{The toy model} \label{sec:appendix_A}

We construct a toy model to quantitatively examine the scenario proposed in
Section~\ref{sec:result:merger} for the baryonic effect on accelerating the
merger process of two galaxy clusters (see Section~\ref{sec:result:toymodel}).
In this toy model, the merger between two galaxy clusters is modeled by the
motion of two point masses, each representing one of the two progenitor
galaxy clusters. Each particle is exerted by three
different forces, i.e., the gravitational force due to the other particle
(or the corresponding DM halo), the dynamical friction due to its relative
motion in the other DM halo, and the additional gravitational drag contributed
by the baryons (see Eq.~\ref{eq:force}). The specific implementation of each
term in the calculation is given as follows.

\begin{figure*}
    \centering
    \includegraphics[width=0.85\textwidth]{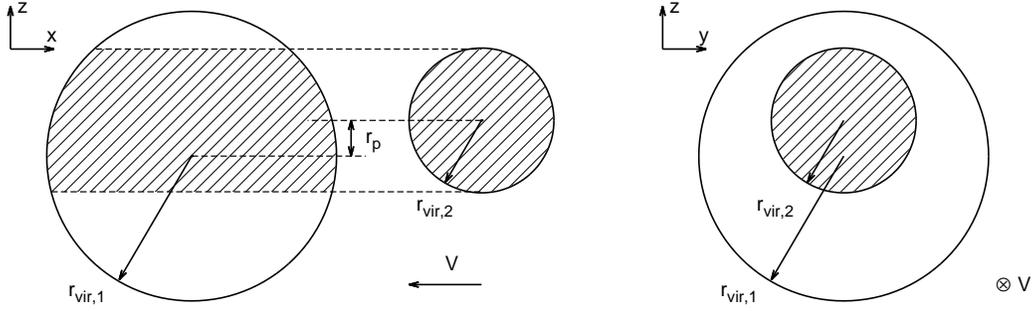}
    \caption{A schematic diagram of the geometric factor $A_{\rm G}(r_{\rm p})$ in
Equation~(\ref{eq:geofac}). The factor $A_{\rm G}(r_{\rm p})$ is defined as the ratio
of the shaded volume to the total volume of the two clusters, where
$r_{\rm p}$ is the pericenter distance for the Keplerian motion of the two-body
system. The virial radii of the two clusters are $r_{\rm vir,1}$ and $r_{\rm
vir,2}$, respectively. The amount of gas in the shaded regions is assumed to be
impeded to be around the center of mass of the entire system after the primary
pericentric passage.}
    \label{fig:diagram}
    \end{figure*}

\begin{figure}[!h]
\centering
\includegraphics[width=0.45\textwidth]{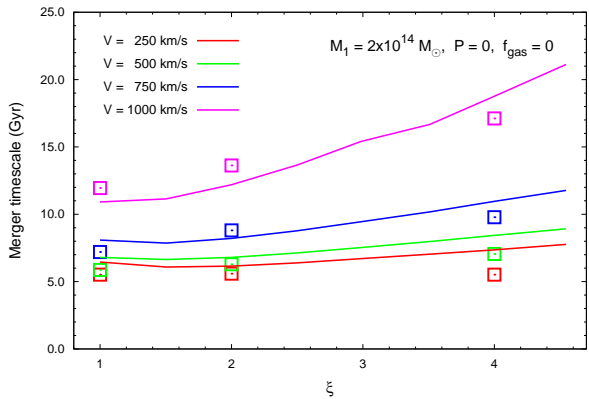}
\caption{Same as in Figure~\ref{fig:parameter}(a), but for different relative
velocities with $(M_1,\,P,\,f_{\rm gas})=(2\times10^{14}\msun,\,0,\,0)$.}
\label{fig:model}
\end{figure}

\begin{itemize}
  \item The acceleration of the particle $i\ (=1,\,2)$ given by the gravitational attraction
  from the DM halo mass of the other particle $j\ (=1,\,2;\,j\neq i$) is,
      \be \cfrac{{\rm d}\textbf{v}_i}{{\rm d}t}\Big|_{\rm halo} = \frac{\textbf{F}_{{\rm halo},\,i}}{M_i} =
      -\frac{G(1-f_{\rm gas})M_j}{r_{ij}^3}\,\textbf{r}_{ij},
      \label{eq:gravforce}
      \ee
      where $\textbf{r}_{ij}=\textbf{r}_{i}-\textbf{r}_{j}$, $r_{ij}\equiv|\textbf{r}_{ij}|$ is the distance between the two particles, $\textbf{r}_{i}$ and $\textbf{v}_{i}$ represent the
      position and the velocity vectors of the particle $i$, respectively, and $G$
      is the gravitational constant.
  \item The dynamical friction is given by the Chandrasekhar formula (see Equation~(8.7) in \citealt{Binney2008})
  and the deceleration due to this dynamical friction is given by
      \begin{eqnarray}
      \cfrac{{\rm d}\textbf{v}_i}{{\rm d}t}\Big|_{\rm df} = \frac{\textbf{F}_{{\rm df},\,i}}{M_i}
      &=& -\alpha_{\rm df}\cdot\cfrac{4\pi G^2f_iM_i\rho_j(r_{ij})\ln \Lambda}{v_{ij}^3} \nonumber \\
      & & \times\left[{\rm erf}(X_j)-\cfrac{2X_j}{\sqrt{\pi}}\e^{-X_j^2}\right]\,\textbf{v}_{ij},
      \label{eq:dynfriction}
      \end{eqnarray}
      where $f_i\equiv(1-f_{\rm gas})\min(M_i,\,M_j)/M_i$ is approximately the
      mass fraction of the particle $i$ traveling through the other cluster $j$,
      $\rm erf$ is the error function, and $X_j\equiv v_{ij}/\sqrt{2}\sigma_j$, $\textbf{v}_{ij}=\textbf{v}_{i}-\textbf{v}_{j}$, $v_{ij}\equiv|\textbf{v}_{ij}|$
      and $\sigma_j$ are the relative velocity of the two clusters and the velocity
      dispersion of the cluster $j$, respectively. The Coulomb logarithm in
      Equation~(\ref{eq:dynfriction}) is set as $\Lambda=1+\xi$
      \citep{Springel2001,Kang2005,Kolchin2008}.  The mass density $\rho_j(r_{ij})$
      of the DM halo $j$ is assumed to follow the truncated NFW profile \citep{Navarro1997},
      \begin{eqnarray}
      & \rho_j & (x \equiv r_{ij}/r_{{\rm s},\,j}) = \nonumber \\
      & & \begin{cases}
      \cfrac{\rho_{{\rm s},\,j}}{x(1+x)^2} {\quad \rm if\ } x \leq c_j, \\
      \cfrac{\rho_{{\rm s},\,j}}{c_j(1+c_j)^2}\exp\left(-\cfrac{x-c_j}{x_{{\rm d},j}}\right) {\quad \rm if\ } x > c_j,
      \end{cases}
      \end{eqnarray}
      where $\rho_{{\rm s},\,j},\ r_{{\rm s},\,j},\ c_j(\equiv r_{{\rm vir},\,j}/r_{{\rm s},\,j}),\ r_{{\rm vir},\,j}$
      are the scale density, the scale radius, the concentration parameter, and the virial radius of the cluster $j$,
      respectively. The truncation scale $x_{{\rm d},j}$ is defined as $2(r_{{\rm vir},\,i}+r_{{\rm vir},\,j})/r_{{\rm s},\,j}$
      in this work. We introduce a free parameter $\alpha_{\rm df}$ to regulate the strength of the dynamical friction,
      which is calibrated by the DM-only simulations. We find that $\alpha_{\rm df}=0.37$ gives a good match between
      the model and the simulations (see Figure~\ref{fig:model} as an example).
  \item The additional gas drag is assumed to be the gravitational force from the gas left around
  the center of mass of the system ($\textbf{r}_{c}$) after the primary pericentric passage of the
  two merging clusters. In the toy model, the force from the left gas halo is switched on at the
  primary pericentric passage, when the center of the gas halo is assumed to be located at the
  center of mass of the system and have the same velocity as the mass center.
  The acceleration of the particle $i$ contributed by this drag can be written as,
    \be
      \cfrac{{\rm d}\textbf{v}_i}{{\rm d}t}\Big|_{\rm gd} =
      \frac{\textbf{F}_{{\rm gd},\,i}}{M_i}=-\alpha_{\rm gd}\cdot\frac{GM_{\rm gas}(r_{ic})}{r_{ic}^3}\,\textbf{r}_{ic},
      \label{eq:gasdrag}
    \ee
    where $\textbf{r}_{ic}\equiv\textbf{r}_{i}-\textbf{r}_{c}$, $r_{ic}\equiv|\textbf{r}_{i}-\textbf{r}_{c}|$
    is the distance between the position of the particle $i$ and the center of the gas halo,
    and $M_{\rm gas}(r_i)$ represents the gas mass within the radius $r_i$.
    The gas density distribution $\rho_{\rm gas}(r)$ is assumed to follow the Hernquist
    model truncated at the radius $r_{\rm b}(\equiv r_{\rm vir,2})$,
      \begin{eqnarray}
      \rho_{\rm gas}(r) = \nonumber
      \begin{cases}
      \cfrac{\rho_{\rm g0}}{(r/a_s)(1+r/a_s)^3} & {\quad \rm if\ } r \leq r_{\rm b}, \\
      0 & {\quad \rm if\ } r > r_{\rm b},
      \end{cases}
      \end{eqnarray}
      where $\rho_{\rm g0}$ and $a_s(\equiv 0.1r_{\rm b})$ is the scale density and scale radius, respectively. The distribution is normalized by the gas halo mass within $r_{\rm b}$, i.e.,
    \begin{eqnarray}
      A_{\rm G}(r_{\rm p})\cdot f_{\rm gas}\cdot (M_1 & + & M_2) = \nonumber \\
      && 4\pi \int_{0}^{r_{\rm b}}\rho_{\rm gas}(r)r^2{\rm d}r,
    \label{eq:geofac}
    \end{eqnarray}
    where $A_{\rm G}(r_{\rm p})$ is a geometric factor as a function of the pericenter
distance $r_{\rm p}$. We estimate $r_{\rm p}$ analytically from the
Keplerian motion of the two-body system by using Equations~(3)--(6) in \citet{Khochfar2006}.
Figure~\ref{fig:diagram} shows the schematic diagram for the definition of
the geometric factor $A_{\rm G}(r_{\rm p})$, which is the ratio of the shaded volume to the
total volume of the two clusters. The factor is assumed to be the fraction of
the gas mass impeded in the center of the merger system after the primary
pericentric passage.  The free parameter $\alpha_{\rm gd}=0.12$
gives a good match between the model and the simulations (see Figure~\ref{fig:parameter}).
\end{itemize}

We solve Equation~(\ref{eq:force}) by the fourth-order Runge-Kutta method.  We
also use the Kustaanheimo-Stiefel regularization to avoid the singularity in
the numerical calculations (see chap.~2 in \citealt{Stiefel1975} for more
details).  The relaxation moment of the merger in the toy model is defined as
the moment of the $n$th pericentric passage, if the time duration between the
$n$th and the $(n+1)$th pericentric passages is smaller than $1/10$ of the time
duration between the primary and the secondary pericentric passages. We find
$n$ ranges between 3--6 in most of the models, roughly consistent with the
results from the numerical simulations.

Figure~\ref{fig:model} shows the merger timescales obtained from both the toy
model and the DM-only numerical simulations for mergers with different mass
ratios and initial relative velocities. As seen from this figure, the merger
timescales resulting from the toy model are well consistent with those obtained
from the numerical simulations.  The gravity between the clusters and the
dynamical friction (i.e., the first two terms in Eq.~\ref{eq:force}) could
model the merger timescale well for the DM-only mergers without baryons, but not the dependence
of the merger timescale on gas fractions for mergers with baryons.  This
implies the importance of the additional gas drag (i.e., the third term in
Eq.~\ref{eq:force}) in modeling the baryonic effect on the merger timescale for
the galaxy clusters.

\end{document}